\documentclass{elsart}
\begin{document}
\begin{frontmatter}
\title{Crossing Probabilities in Critical 2-D Percolation and Modular Forms
}

\author{Peter Kleban \thanksref{em}}
\address{LASST \\
and \\
Department of Physics and Astronomy \\
University of Maine \\
Orono, Maine 04469 \\
USA}
\thanks[em]{Email: email: kleban@maine.edu
; tel: (207) 581-2258
; fax: (207) 581-2255}

\maketitle

\begin{abstract}

Crossing probabilities for critical 2-D percolation on large but finite
lattices have been derived via boundary conformal field theory.
These predictions agree very well with numerical results.
However, their derivation is heuristic and there is evidence of
additional symmetries in the problem.
This contribution gives a preliminary examination some unusual modular
behavior of these quantities. In particular, the derivatives of
the "horizontal" and "horizontal-vertical" crossing probabilities
transform as a vector modular form, one component of which is an
ordinary modular form and the other the product of a modular form
with the integral of a modular form. We include consideration of the
interplay between conformal and modular invariance that arises.


\end{abstract}
\end{frontmatter}

\section{Introduction}
Percolation is perhaps the simplest non-trivial model in statistical
mechanics. It is very easy to define, and exhibits a second-order phase
transition between the percolating and non-percolating states. A broad
array of techniques have been brought to bear on it over a period of many
years. Its behavior is of current interest, and has been studied via
renormalization group, conformal field theory, Coulomb gas methods,
computer simulation, as an example of supersymmetry, and using rigorous
mathematical methods. (For general reviews, see [1],[2], for some recent
results including a list of references see [3]). In this contribution, we
restrict ourselves to the study of crossing probabilities for
two-dimensional systems at the percolation (phase transition) point $p_c$.

Although percolation is, as mentioned, arguably the simplest model that
exhibits a second order phase transition, the ease of formulation of
the model is in another sense deceptive, tending to conceal its
inherent complexity. The wide range of approaches taken to it already
attests to this subtlety. The ultimate reason is suspected to be the
unconstrained nature of the model, encompassing a variety of symmetries.

In Section 2 we review percolation and crossing probabilities.
Exact analytic expressions for these quantities are known from
boundary conformal field theory. We transform these results into a
form suitable for the present analysis. Section 3 briefly introduces
modular forms. Then the unusual modular behavior of the derivatives
of the crossing probabilities is examined. The results here are
preliminary; a full treatment will appear elsewhere [4].

\section{Crossing Probabilities}

The properties that we consider here are critical and therefore
universal, the same for a wide variety of types of (isotropic)
percolation and lattice structures. However, for definiteness, when
we are specific we will refer to bond percolation on a square lattice.
Bond percolation is defined by placing a bond with (independent)
probability $p$ on each edge of the lattice.
Consequently, there are $2^N$ possible bond configurations with $0 \leq
N_B
\leq N$,
where $N_B$ is the number of bonds in a given configuration and $N$ is
the total number of edges. The connected bonds in each configuration form
clusters. At $p_c$, (note that duality implies $p_c = 1/2$ on a square
lattice),
as the lattice is taken to infinity, one or more infinite clusters just
appears.  For $p < p_c$, there is no infinite cluster.

The crossing probabilities are defined by considering a finite
rectangular $L \times L'$ lattice as $L, L' \rightarrow$ infinity
with fixed aspect
ratio $r$ = width/height = $L/L'$. (Below, we will allow the shape
of the lattice to change.) Then the probability of a configuration
connecting the left side and the right side of the rectangle is
the horizontal crossing probability $\Pi_h(r)$. The probability of a
configuration connecting all four sides is the horizontal-vertical
crossing probability $\Pi_{hv}(r)$. These quantities are known to depend
only on the aspect ratio $r$ (for a rectangle) by extensive numerical
work and the hypothesis (and consequences) of conformal invariance.
In fact, they enjoy an even wider invariance, as discussed below.

Next, to motivate the conformal approach to this problem,
we consider the $Q$-state Potts model. This generalization of the
Ising model employs a spin variable $s_i$ with values $s_i = 1, 2, ..., Q$
($Q = 2$ corresponds to the Ising model) on each site $i$ of the lattice.
The (reduced) Hamiltonian is

\begin{equation}
H=K \sum _{<ij>} \delta _{s_i,s_j}
\end{equation}

where $<ij>$ denotes nearest neighbor sites.
By introducing the variable $x = e^K - 1$ one may rewrite the partition
function as follows:

\begin{equation}
Z= \sum _{\{s_i\}} \prod_{<i,j>} (1+x \delta_{s_i,s_j})
\end{equation}

Expanding the product, one may perform the sum over $s_i$ in each term.
Representing the presence of a factor $x$ by a bond then gives rise to
a graphical representation of $Z$, known as the random cluster or
bond-correlated Potts representation [5],[6-9]

\begin{equation}
Z= \sum_{G} Q^{N_c} x^{N_b}
\end{equation}

where the sum is over all possible graphs consisting of $N_b$
bonds arranged in $N_c$ clusters
(counting single isolated sites as clusters).
This model is known to have a critical point for all $0 \leq Q \leq 4$.
(On the square lattice, by duality, $x_c = Q^{1/2}$).
For $Q = 1$, the set of configurations included is exactly that,
including the weighting, as for bond percolation, with $p = x/(x+1)$.
For other $Q$ values, the configurations are the same but weighted
differently. In addition, Eq. (3) allows us to extend the number of
states to $Q \in {\bf R}$. Thus we can envision a continuous change from
the
Ising model, say, to percolation. Further, the central charge is
known as a function of $Q$; in particular, $c = 0$ for $Q = 1$
(critical percolation).

In order to study the crossing probabilities, following [10],
consider for definiteness $\Pi_h$ on a rectangle.
Let $Z_{ab}$ be the partition function of the Potts model with the
spins on the left vertical side fixed in state $a$ and those on the
right vertical side fixed in state $b$. The spins on the rest of
the boundary are unrestricted. Then

\begin{equation}
\Pi_{h} = \lim_{Q \to 1} \left( 1- \frac{Z_{ab}}{Z_{aa}} \right)
\end{equation}

with $a \neq b$. Note that this expression differs from the one in [10]
due to the normalization of $Z$. The expression makes no sense for
$Q = 1$, of course, but it allows a solution to the problem if we
first express the partition functions using boundary operators and
then take the limit.

Since we are at criticality, the partition function can
(in the limit of a large lattice) be expressed using conformal
field theory. The critical Potts models are known to correspond
to a certain series of minimal models. A change of boundary
conditions is introduced by means of a boundary operator [11].
If the coordinates of the corners of the rectangle are $z_1, z_2, z_3,
z_4$
we thus have

\begin{equation}
Z_{ab}= Z_f \left< \phi_{fa} ( z_1) \phi_{af} (z_2) \phi_{fb} (z_3)
\phi_{bf}
(z_4) \right>
\end{equation}

where $Z_f$ is the partition function with free boundary conditions
and the $\phi$s  are boundary operators. The next step is to identify
 $\phi_{af}$.
This is done by comparison with known results for the Ising and $Q = 3$
state Potts models. In these cases, the operator that changes between
fixed boundary conditions $a$ and $b$ is known to be $\phi_{(1,3)}$. On
the other
hand, one can implement this change by bringing together two points $z_1$
and $z_2$ where the boundary conditions go from $a$ to $f$ and $f$ to $b$,
respectively. Using the operator product expansion then gives a term
that must be the operator in question. By the fusion rules, the only
operator that can satisfy this is seen to be $\phi_{af} = \phi_{(1,2)}$ .
This argument is doubly satisfying, since the conformal dimension
of $\phi_{(1,2)}$ is $h = 0$ in the limit $Q \rightarrow 1$, which is a
necessary
requirement for the crossing probability to be conformally invariant.
Further, the operator is level two, so that the differential equation
satisfied by its four-point function is second order.

It is conventional to consider the problem on the upper half plane,
which may subsequently be mapped onto a rectangle via the
Schwarz-Christoffel transformation, with the four corner points
taken as images of $-1/k, -1, 1, 1/k$ (with $0 < k < 1$).
Then the crossing is between the intervals $-1/k < x < -1$ and
$1 < x < 1/k$ on the real axis. The four-point correlation
functions depend only on the cross-ratio

\begin{equation}
\lambda = \frac{(x_4 - x_3)(x_2 - x_1)}{(x_3 -x_1)(x_4 -x_2)}=
\left( \frac{1-k}{1+k} \right) ^2
\end{equation}

One then finds, by means of standard conformal manipulations,
that the correlation function satisfies a Riemann equation
with the two solutions $F(\lambda) = 1$ and $F(\lambda) = \lambda^{1/3}
{}_2 F_1(1/3,2/3;4/3;\lambda)$.
One can pick the correct linear combination by imposing the physical
constraints that $F \to 0$ as $\lambda \to 0$ ($r \to \infty$ )
and $F \to 1$ as $\lambda \to 1$ ($r \to 0$).
The result is

\begin{equation}
\Pi_h (\lambda) = \frac{2 \pi \sqrt{3}}{\Gamma (1/3)^3} \lambda^{1/3}
{}_2 F_1(1/3,2/3;4/3;\lambda)
\end{equation}

Mapping Eq. (7) onto a rectangle, one finds that the aspect
ratio becomes $r = 2K/K'$, where $K'$ and $K$ are the complete elliptic
integrals. This result has been extensively tested via Monte Carlo
simulations [12],[13], and there is little doubt as to its correctness.
Note that one can transform it to any compact shape with four
identified points, not just a rectangle; the crossing probability will
be the same as the crossing on the half-plane with the corresponding
half-plane cross-ratio. Thus, for instance, by consideration of the
Schwarz-Christoffel formula it is easy to see that the crossing
probability on a rhombus of any angle is the same as on a square.
This point has been investigated numerically in Fig. 4.1 of [13].
Of course the same invariance holds for the "horizontal-vertical"
crossing considered below, only the function $F$ changes.

In general, when one makes a conformal transformation $z \to w(z)$ of
a correlation function, factors of $(w'(z))^h$ appear. In addition,
transforming from the upper half plane to a shape with corners,
the partition function gains a (non-scale invariant) factor $L^{ac}$,
where $L$ is the length scale, $c$ is the central charge and $a$ depends
on the geometry. Similarly, a correlation function with boundary
operators sitting at the corner points gains a factor
$L^{-(\pi/\gamma)h}$,
where $\gamma$ is the interior angle at the corner [14]. The last two
(non-scale invariant) effects occur because the transformation is only
piecewise analytic and has singular points at the corners. However,
for critical percolation, $c = 0 = h$, and all of these factors become
unity. Thus the crossing probabilities are invariant under
transformations that are conformally invariant in the interior of a
region and only piecewise conformally invariant on its edges.

The "horizontal-vertical" crossing probability $\Pi_{hv}$ may be obtained
similarly [15]. We omit most details of the argument.
A four-point boundary operator correlation function arises once again.
The main differences with $\Pi_h$ are the complexity of the
boundary conditions and the absence of any direct identification
of the boundary operator. Instead, one considers
low-lying null vectors in the $c = 0 = h$ Verma module.
Through level 5, there is only one which leads to solutions
that satisfy the physical requirements. These are
\begin{eqnarray}
\Pi_{hv}(r) & = & \Pi_{hv}(1/r) \nonumber \\
\Pi_{hv}(r) & \stackrel{r \to \infty}{\longrightarrow} & \Pi_h (r)
\end{eqnarray}
which translate into
\begin{eqnarray}
F(\lambda)&=&F(1-\lambda) \nonumber \\
F(\lambda) &\stackrel{\lambda \to 0}{\longrightarrow} & \Pi_h(\lambda)
\end{eqnarray}
respectively. Applying these conditions to the level 5 solutions,
one finds

\begin{equation}
\Pi_{hv}(\lambda)=\frac{2 \pi \sqrt{3}}{\Gamma (1/3)^{3}} \lambda^{1/3}
{}_2 F_1(1/3,2/3;4/3;\lambda)-\frac{\sqrt{3}}{2 \pi} \lambda {}_3
F_2(1,1,4/3
;2,5/3;\lambda)
\end{equation}

where ${}_3F_2$ is a generalized hypergeometric function.
The first term is just $\Pi_h$ and the second subtracts configurations
with
horizontal but no vertical crossings.

The differential equation satisfied by $F$ may be written
\begin{eqnarray}
 && \frac{d^3}{d\lambda^3}( \lambda (\lambda-
1))^{4/3}\frac{d}{d\lambda}(\lambda(
\lambda-1))^{2/3}\frac{d}{d\lambda}F
=    \nonumber \\
&& \left[ \frac{d^2}{d\lambda^2}(\lambda(\lambda-1))+\frac{1}{2\lambda-1}
\frac{d}{d\lambda}(2\lambda-1)^{2} \right]
\left[\frac{d}{d\lambda}(\lambda(\lambda-1))^{1/3}\frac{d}{d\lambda}
(\lambda(\lambda-1))^{2/3}\frac{d}{d\lambda} \right] F
=0    \nonumber \\
&&
\end{eqnarray}

The factorized form exhibited in the second line is of interest
since 1, $\Pi_h$, and $\Pi_{hv}$
span the solutions of the equation formed by
letting the rightmost factor act on $F$ alone, i.e.

\begin{equation}
\left[ \frac{d}{d \lambda} (\lambda(\lambda-1))^{1/3} \frac{d}{d\lambda}
(\lambda(\lambda-1))^{2/3} \frac{d}{d \lambda} \right] F = 0
\end{equation}

In what follows, it is convenient to consider the $r$-derivatives
of $\Pi_h$ and $\Pi_{hv}$ (on the rectangle) which we will denote
$\Pi_h'(r)$
and $\Pi_{hv}'(r)$. Note that $\Pi_h'$, for instance, is interpretable
physically as the probability density that the maximum horizontal
extent of a cluster attached to one vertical side of an
infinitely wide rectangle of unit height is greater than $r$ [16].
Additionally, since the $r$-derivative is proportional to the
$\lambda$-derivative, Eq. (12) reduces to second order.

We next proceed to express $\Pi_h'$ and $\Pi_{hv}'$ on the rectangle as
functions of $r$, using the result for the cross-ratio [16, 17]

\begin{equation}
\lambda= \left( \frac{\vartheta_2({\hat q})}{\vartheta_3({\hat q})}
\right) ^4
\end{equation}

where $\vartheta_2$ and $\vartheta_3$ are the elliptic theta-functions and
${\hat q}=e^{-\pi r}$
(note that ${\hat q}$ is the square root of the usual $q$). Eq. (13)
follows by applying Landen's transformation to $r = 2K/K'$
and Eq. (6), resulting in

\begin{equation}
r= \frac{K (\sqrt{1-\lambda})}{K(\sqrt{\lambda})}
\end{equation}

where $K$ is the complete elliptic integral written as a
function of the modulus. Expressing the latter in terms of a ratio
of theta-functions ([18] 8.197.1,2) one obtains Eq. (13).

We also note, for future reference, some identities involving $\lambda$.
One has $\lambda=16 \frac{\eta(\tau/2)^8 \eta(2
\tau)^{16}}{\eta(\tau)^{24}}$,
$1-\lambda= \frac{\eta(\tau/2)^{16} \eta(2 \tau)^{8}}{\eta(\tau)^{24}}$,
and $\lambda'=16 \frac{\eta(\tau/2)^{16} \eta(2
\tau)^{16}}{\eta(\tau)^{28}}$,
where $\eta$ is the Dedekind $\eta$-function (see Eq. (18))
with $q = e^{2\pi i\tau}$, and the differentiation is with respect to the
independent variable $\tau = ir$.

Eq. (13) makes it possible to re-write the differential Eq. (12)
for $F$ directly in terms of the aspect ratio $r$. One obtains

\begin{equation}
\frac{d^2 f}{d r^2}+a(r)\frac{d f}{d r}+b(r)f=0
\end{equation}

where $f(r) \equiv F'(r)$, the $r$-derivative of the four-point function,
and

$$
a(r)=-\frac{3 \lambda ''}{\lambda'}+\frac{5 \lambda '}{3(\lambda-1)}
+\frac{5 \lambda '}{3 \lambda}
$$
\begin{equation}
b(r)=-\frac{5 \lambda ''}{3 \lambda}+3 \left( \frac{\lambda ''}{\lambda'}
\right)^2
-\frac{\lambda '''}{\lambda'}-\frac{5 \lambda ''}{3(\lambda-1)}
+\frac{4 (\lambda ')^2}{3 \lambda(\lambda-1)}
\end{equation}

where differentiation is with respect to $r$. One may identify two
independent solutions of Eq. (15) as

$$
f_1=\left[ \eta({\hat q}^2)\right]^4
$$
\begin{equation}
f_2=\frac{1}{2}\left[\vartheta_2({\hat q})\right]^4-f_W
\end{equation}

where $\eta$ is the Dedekind $\eta$-function

\begin{equation}
\eta(q)=q^{1/24}\prod^{\infty}_{n=1}(1-q^n)
\end{equation}

and $f_W$ is an even, and apparently new, function of ${\hat q}$.
Its first few terms are given by

\begin{equation}
f_W = \frac{16}{5}({\hat q}^2+\frac{16}{11}{\hat q}^4+ \frac{364}{187}
{\hat q}^6
+ \frac{13568}{4301} {\hat q}^8 + \frac{458070}{124729} {\hat q}^{10} +
...)
\end{equation}

The solution $f_1$ is seen to be proportional to $\Pi_h'$ by the above, or
directly [17]. We note for completeness that

\begin{equation}
\Pi_{h}'(r)=- \frac{2^{7/3} \pi^2}{\sqrt{3} \Gamma (1/3)^3}
\left[ \eta({\hat q}^2) \right] ^4
\end{equation}

The connection with $\Pi_{hv}'$ involves including
a term proportional to $f_2$.
This is specified below. Note that, because of the way we have
defined the aspect ratio $r$, our $\Pi_h$ coincides with
$\Pi_v$ in [16, 17].

\section{Modular Forms}

This section follows the excellent introduction given in [19].

A modular function or form assigns a complex number $G(\Lambda)$ to each
lattice $\Lambda$. Here the term 'lattice', following mathematical usage,
refers to an infinite regular array of points, defined by the basis
$\{\omega_1,\omega_2 \}: \Lambda = {\bf Z} \omega_1 + {\bf Z} \omega_2$.
In addition, for any complex number $\lambda \neq 0$, $G$
satisfies $G(\Lambda) = \lambda^k G(\lambda \Lambda)$,
where $k$ is some integer, called the weight
(with $k = 0$ for modular functions). Because of this, on dividing by
$\omega_2$,
we see that $G$ is completely specified by $g(\tau) = G({\bf Z}\tau + {\bf
Z})$,
where $\tau = \omega_1/\omega_2$. Since $g$ is also even in $\tau$, we may
restrict
$\tau$ to
the upper half plane. In addition, there are analyticity and growth
(as $\tau \to i \infty$) conditions on $g$.

The modular properties arise on considering a change of basis.
We can replace $\{\omega_1,\omega_2\}$ by $\{\omega_1',\omega_2'\}$ =
$\{a\omega_1 + b\omega_2,c\omega_1 + d\omega_2\}$ with
$a,b,c,d \in {\bf Z}$,
$ad - bc = \pm 1$ without changing the lattice.
This implies that $g$ must satisfy the modular transformation property

\begin{equation}
g \left( \frac{a\tau +b}{c\tau +d} \right) = (c \tau + d)^k g(\tau)
\end{equation}

Restricting $\tau$ to the upper half plane imposes $ad - bc = +1$.
The group of matrices implementing such transformations is the (full)
modular group $\Gamma_1$. The matrix
$T = \left( \begin{array}{cc}
1 & 1 \\
0 & 1 \end{array}
\right)$
, which implements $\tau \to \tau + 1$, together
with
$S = \left( \begin{array}{cc}
0 & -1 \\
1 & 0 \end{array}
\right)$
, for $\tau \to -1/\tau$, generate $\Gamma_1$.

It follows from Eq. (21) that modular forms of a given weight are a
vector space over ${\bf C}$. The dimension of this space is very small,
in general, a fact which leads to some very non-trivial relations.

In what follows, we examine the modular properties of $f_1$ and $f_2$,
considered as functions of $\tau = ir$, with $r$ complex.
(Here it is useful to envision the rectangle rotated by $90^{\circ}$,
since $\tau$ is in the upper half plane for real $r$.)
In order to make sense of this, we need to understand the
physical meaning of a modular transformation in the context of
percolation crossing probabilities. We do not include a full
treatment of this question here, but restrict ourselves to the
case of a conformally invariant physical quantity $\Pi$,
which is also a modular function (i.e. invariant under any
$\gamma \in \Gamma_1$).
Suppose further that $\Pi$ is initially defined on a rectangle of
aspect ratio $r$. Now the modular transformation, as indicated,
acts only on the basis vectors $\{\omega_1,\omega_2\}$.
(For a rectangle, Re$\{\omega_1\} = 0 =$ Im $\{\omega_2\}$). To construct
a conformal
map of the rectangle, we consider these as displacements from a
fixed origin at 0. Thus the map must satisfy $\{0,\omega_1,\omega_2\} \to
\{0,\omega_1',\omega_2'\} =
\{0,a\omega_1 + b\omega_2,c\omega_1 + d\omega_2\}$. This can be
implemented, for instance,
by a projective transformation

$$
w = \frac{\alpha z}{\epsilon z+\delta},
$$
$$
\alpha = (\omega_2 -\omega_1)
 \omega_1'
\omega _2',
$$
$$
\epsilon = \omega _1'  \omega _2
- \omega_1 \omega _2',
$$
\begin{equation}
\delta = \omega_1 \omega_2 ( \omega_2'
- \omega_1')
\end{equation}

which will take the rectangle into a figure with sides that
are straight or arcs of circles. Note that the necessary conditions
$\omega_1 \neq \omega_2$ and $\omega_1'\neq \omega_2'$ imply
$\alpha, \delta \neq 0$.
This choice of map also
has the advantage of always being 1-to-1.  In addition, one can show that
it preserves the group structure.  However, the map is explicitly
dependent on the basis vectors. Note that the quantity $\Pi$ will
remain invariant. The parameter $\tau$, defined as a ratio of
displacements, transforms in the usual way, so that
$\Pi(\gamma(\tau)) = \Pi(\tau)$.
Thus for each $\gamma$, modular invariance corresponds to conformal
invariance under a particular map. Of course we can also begin
with the physical quantity defined on a parallelogram,
removing the conditions on $\{\omega_1,\omega_2\}$. Note, however,
that the derivative of $w$ at either of the three corners
differs from the derivative of the modular transformation.

We now return to the crossing probabilities.
We focus on the functions $f_1$ and $f_2$, solutions of Eq. (15),
considering them as functions of $\tau = ir$, so that
${\hat q}=e^{\pi i \tau}$.
We have already specified the connection between $f_1$ and $\Pi_h'$.
As mentioned, the solutions of Eq. (12)
(and therefore those of Eq. (15)) span $\Pi_h'$ and $\Pi_{hv}'$.
Thus it only remains to find the correct linear combination of $f_1$ and
$f_2$ . Now $\Pi_{hv}'$ must, by the first of Eqs. (9),
be invariant under $W: f(\tau) \to \tau^{-2} f(-1/\tau)$.
The linear combination $f_2-\frac{C}{2} f_1$ with $C=\frac{2^{1/3} \pi ^2}
{3 \Gamma (1/3)^3}$
satisfies this condition.
Combining this with Eq. (20) we find that

\begin{equation}
\Pi_{hv}'(r) = - \frac{2^{7/3} \pi ^2}{\sqrt{3} \Gamma (1/3)^3}
\left[ \eta({\hat q}^2) \right] ^4 + \frac{24}{\sqrt{3}} f_2 ({\hat q})
\end{equation}

Now this function behaves like a weight 2 modular form under
the operations $\tau \to -1/\tau$ and $\tau \to \tau + 6$,
 but these transformations
generate a subgroup of infinite index of $SL_2({\bf Z})$, and
$\Pi_{hv}'$ is
not a modular form. However, the vector
$F = \left( \begin{array}{c}
f_1 \\
f_2 \end{array}
\right)$
transforms under $T^2$ and $S$,
which generate the theta-group $\Gamma_\theta$, according to

$$ F(\tau +2)= \left( \begin{array}{cc}
\omega & 0 \\
0 & 1 \end{array}
\right)F(\tau)
$$
\begin{equation}
\tau ^{-2} F(-1/\tau) = \left( \begin{array}{cc}
-1 & 0 \\
-C & 1 \end{array}
\right)
F(\tau)
\end{equation}

where $C$ is as above and $\omega = e^{2\pi i/3}$. Thus $f_1$ is a weight
2 modular
form (with multiplier system, cf. [19]) on $\Gamma_\theta$,
but $f_2$ is a kind of
"second order" modular form, i.e. instead of $\gamma|f_2 - f_2 = 0$ for any
$\gamma \in \Gamma_\theta$,
(where $|$ denotes the modular weight 2 operation of $2 \times 2$
matrices)
we have that $\gamma|f_2 - f_2$ is a multiple of $f_1$.

Another way to look at this situation is to consider the
differential equation (15). For convenience we take the independent
variable to be $\tau = ir$. Then, writing $f_2 = uf_1$, we find a
second-order equation $f_1 u'' + (2f_1'+af_1) u'=0$
for $u'$. Now $a = -h'/h$, where $h={\lambda'}^3 \lambda^{-5/3}
(1-\lambda)^{-5/3}$
 so that $u''/u'=-a-2f_1'/f_1$.
Hence $u'=2/3 h f_1^{-2}$.
Using Eq. (17) for $f_1$ and the identities for $\lambda$ above, we have
finally

\begin{equation}
u'(\tau)=\frac{2}{3} \frac{\eta (\tau/2)^8 \eta (2 \tau)^8}{\eta
(\tau)^{12}}
\end{equation}

It follows from the modular properties of $\eta$ that $u'$ is a modular
form of weight 2 on $\Gamma_\theta$,
so that $f_2$ is the product of a modular form
and the integral of a modular form.

Our examination of the modular properties of $f_1$ and $f_2$ reveals a
close connection between the two functions. Now modular forms, as
mentioned, are vector spaces of very small dimension. Thus it is
likely that one can derive $f_2$ given $f_1$ and some physical conditions.
This would be very interesting. Whether it is indeed possible remains
to be seen, however.  We have succeeded in showing that $f_1$ follows (up
to an overall multiplicative constant) if one assumes that only one
conformal block contributes to it [4].

One can also show that the function $u$ corresponds to the Weierstrass
$\zeta$-function on the elliptic curve $Y^2 = X^3 -1728$.

\section*{Acknowledgements}

We acknowledge useful conversations with D. Bradley, J. Cardy, A.
\"{O}zl\"{u}k,
I. Peschel, C. Snyder, D. Zagier and R. Ziff.

\end{document}